\documentclass[3p,times,procedia]{elsarticle}

\usepackage{nuphb_ecrc}
\usepackage{amsmath}

\volume{00}
\firstpage{1}
\journalname{Nuclear Physics B}
\runauth{M.C.~Gonzalez-Garcia et al.}
\jid{nuphb}
\jnltitlelogo{Nuclear Physics B}
\biboptions{sort&compress}

\newcommand{\Dmq}{\Delta m^2}
\newcommand{\eVq}{\ensuremath{\text{eV}^2}}
\newcommand{\Nuc}[2]{\ensuremath{\mbox{}^{#1}\text{#2}}}
\newcommand{\diag}{\mathop{\mathrm{diag}}}
\newcommand{\sign}{\mathop{\mathrm{sign}}}

\renewcommand{\Im}{\mathop{\mathrm{Im}}}
\newcommand{\Eps}{\varepsilon}

\begin{document}

\begin{frontmatter}

\title{Global Analyses of Neutrino Oscillation Experiments}

\author[s111,s112]{M.C. Gonzalez-Garcia}
\address[s111]{Instituci\'o Catalana de Recerca i Estudis Avan\c{c}ats
  (ICREA), Departament d'Estructura i Constituents de la Mat\`eria and
  Institut de Ciencies del Cosmos, Universitat de Barcelona, Diagonal
  647, E-08028 Barcelona, Spain}
\address[s112]{C.N.~Yang Institute for Theoretical Physics, State
  University of New York at Stony Brook, Stony Brook, NY 11794-3840,
  USA}
\ead{maria.gonzalez-garcia@stonybrook.edu}

\author[s113]{Michele Maltoni}
\address[s113]{Instituto de F\'{\i}sica Te\'orica UAM/CSIC, Calle de
  Nicol\'as Cabrera 13--15, Universidad Aut\'onoma de Madrid,
  Cantoblanco, E-28049 Madrid, Spain}
\ead{michele.maltoni@csic.es}

\author[s114]{Thomas Schwetz}
\address[s114]{Institut f\"ur Kernphysik, Karlsruher Institut f\"ur
  Technologie (KIT), D-76021 Karlsruhe, Germany}
\ead{schwetz@kit.edu}

\begin{abstract}
  We summarize the determination of some neutrino properties from the
  global analysis of solar, atmospheric, reactor, and accelerator
  neutrino data in the framework of three-neutrino mixing as well as
  in some extended scenarios such as the mixing with eV-scale sterile
  neutrinos invoked for the interpretation of the short baseline
  anomalies, and the presence of non-standard neutrino interactions.
\end{abstract}

\end{frontmatter}

\section{Introduction: the New Minimal Standard Model}

Thanks to remarkable discoveries by a number of neutrino oscillation
experiments it is now an established fact that neutrinos have mass and
leptonic flavors are not symmetries of Nature~\cite{Pontecorvo:1967fh,
  Gribov:1968kq}. Historically neutrino oscillations were first
observed in the disappearance of solar $\nu_e$'s and atmospheric
$\nu_\mu$'s which could be interpreted as flavor oscillations with two
very different wavelengths. Over the last 15 years, these effects were
confirmed also with terrestrial experiments using man made beams from
accelerators and nuclear reactors (see
Ref.~\cite{GonzalezGarcia:2007ib} for an overview).  In brief, at
present we have observed neutrino oscillation effects in:
\begin{itemize}
\item atmospheric neutrinos, in particular in the high-statistics
  results of Super-Kamiokande~\cite{skatm:nu2014};

\item event rates of solar neutrino radiochemical experiments
  Chlorine~\cite{Cleveland:1998nv}, Gallex/GNO~\cite{Kaether:2010ag}
  and SAGE~\cite{Abdurashitov:2009tn}, as well as time and energy
  dependent rates from the four phases in
  Super-Kamiokande~\cite{Hosaka:2005um, Cravens:2008aa, Abe:2010hy,
    sksol:nu2014}, the three phases of SNO~\cite{Aharmim:2011vm}, and
  Borexino~\cite{Bellini:2011rx, Bellini:2008mr};

\item disappearance results from accelerator long baseline (LBL)
  experiments in the form of the energy distribution of $\nu_\mu$ and
  $\bar\nu_\mu$ events in MINOS~\cite{Adamson:2013whj} and
  T2K~\cite{Abe:2014ugx}, and $\nu_\mu$ events in
  NO$\nu$A~\cite{nova:nufact15};

\item LBL $\nu_e$ appearance results for both neutrino and
  antineutrino events in MINOS~\cite{Adamson:2013ue}, and $\nu_e$
  appearance in NO$\nu$A~\cite{nova:nufact15} and
  T2K~\cite{Abe:2013hdq};

\item reactor $\bar\nu_e$ disappearance at medium baselines in the
  form of the energy distribution of events in Double
  Chooz~\cite{Abe:2012tg}, Daya Bay~\cite{db:nu2014} and
  RENO~\cite{RENO:2015ksa};

\item the energy spectrum of reactor $\bar\nu_e$ disappearance at LBL
  in KamLAND~\cite{Gando:2010aa}.
\end{itemize}
These results imply that \emph{neutrinos are massive} and \emph{there
  is physics beyond the Standard Model} (SM). The fundamental question
arises, what is the underlying theory for neutrino masses.  In this
article, however, we will focus on the more mundane but difficult
approach of the detailed determination of the simplest low energy
parametrization(s) required to describe the bulk of data.

The SM is a gauge theory based on the gauge symmetry $\mathit{SU}(3)_C
\times \mathit{SU}(2)_L \times U(1)_Y$ spontaneously broken to
$\mathit{SU}(3)_C \times U(1)_\mathit{EM}$ by the the vacuum
expectation value of a Higgs doublet field $\phi$.  The SM contains
three fermion generations which reside in chiral representations of
the gauge group. Right-handed fields are included for charged fermions
as they are needed to build the electromagnetic and strong
currents. However, no right-handed neutrinos are included in the model
since neutrinos are neutral and colourless and therefore the
right-handed neutrinos are singlets of the SM group.

In the SM, fermion masses arise from the Yukawa interactions which
couple the right-handed fermion singlets to the left-handed fermion
doublets and the Higgs doublet. After spontaneous electroweak symmetry
breaking these interactions lead to charged fermion masses but leave
the neutrinos massless. No Yukawa interaction can be written that
would give a tree level mass to the neutrino because no right-handed
neutrino field exists in the model.

Furthermore, within the SM $G_\text{SM}^\text{global} = U(1)_B \times
U(1)_e \times U(1)_\mu \times U(1)_\tau$ is an accidental global
symmetry.  Here $U(1)_B$ is the baryon number symmetry, and
$U(1)_{e,\mu,\tau}$ are the three lepton flavor symmetries. Any
neutrino mass term which could be built with the particle content of
the SM would violate the $U(1)_L$ subgroup of
$G_\text{SM}^\text{global}$ and therefore cannot be induced by loop
corrections.  Also, it cannot be induced by non-perturbative
corrections because the $U(1)_{B-L}$ subgroup of
$G_\text{SM}^\text{global}$ is non-anomalous.

It follows then that the SM predicts that neutrinos are
\emph{strictly} massless.  Consequently, there is neither mixing nor
CP violation in the leptonic sector. Clearly this is in contradiction
with the neutrino data summarized above.  So the Standard Model has to
be extended at least to include neutrino masses.  This minimal
extension is what we call the \emph{New Minimal Standard Model}
(NMSM).

The two minimal extensions to give neutrino mass and explain the data
are:
\begin{itemize}
\item to introduce $\nu_R$ and impose total lepton number $(L)$
  conservation. After spontaneous electroweak symmetry breaking we
  have:
  \begin{equation}
    \label{eq:dirac}
    \mathcal{L}_\text{D} = \mathcal{L}_\text{SM}
    - M_\nu \bar \nu_L \nu_R + \text{h.c.}
  \end{equation}
  In this case mass eigenstate neutrinos are Dirac fermions,
  \textit{i.e.}, $\nu^c \neq \nu$;

\item to construct a mass term only with the SM left-handed neutrinos
  by allowing $L$ violation:
  \begin{equation}
    \mathcal{L}_\text{M} = \mathcal{L}_\text{SM}
    - \frac{1}{2} M_\nu \bar \nu_L \nu_L^c + \text{h.c.}
  \end{equation}
  In this case the mass eigenstates are Majorana fermions, $\nu^c =
  \nu$.  Note that the Majorana mass term above breaks the electroweak
  gauge invariance, and therefore spoils the renormalizability of the
  model. In this respect $\mathcal{L}_\text{M}$ can only be understood
  as a low energy limit of a complete theory, whereas
  $\mathcal{L}_\text{D}$ is formally self-consistent.
\end{itemize}
Either way, in the NMSM flavour is mixed in the CC interactions of the
leptons, and a leptonic mixing matrix appears analogous to the CKM
matrix for the quarks.  However the discussion of leptonic mixing is
complicated by two factors. First the number massive neutrinos ($n$)
is unknown, since there are no constraints on the number of
right-handed (SM-singlet) neutrinos. Second, since neutrinos carry
neither color nor electromagnetic charge, they could be Majorana
fermions. As a consequence the number of new parameters in the model
depends on the number of massive neutrino states and on whether they
are Dirac or Majorana particles.

In general, if we denote the neutrino mass eigenstates by $\nu_i$, $i
= 1,2,\ldots,n$, and the charged lepton mass eigenstates by $l_i =
(e,\mu,\tau)$, in the mass basis, leptonic CC interactions are given
by
\begin{equation}
  \label{eq:CClepmas}
  -\mathcal{L}_\text{CC} = \frac{g}{\sqrt{2}} \,
  \bar{l}_{iL} \, \gamma^\mu \, U^{ij}\, \nu_j \, W_\mu^+
  + \text{h.c.}
\end{equation}
Here $U$ is a $3\times n$ matrix which verifies $U U^\dagger =
I_{3\times 3}$ but in general $U^\dagger U \neq I_{n\times n}$. This
is the case, for example, when considering mixing with non-doublet
states, such as discussed in Sec.~\ref{sec:1evsterile}.

In what follows we will review the status of the analysis of the
oscillation neutrino data in different frameworks. In
Sec.~\ref{sec:3numixing} we present the results for the case of
three-neutrino mixing, and in Sec.~\ref{sec:absmass} we discuss the
implications of such results for observables sensitive to the absolute
neutrino mass scale. In Sec.~\ref{sec:1evsterile} we focus on extended
scenarios involving mixing with eV-scale sterile neutrinos, as invoked
for the interpretation of the short baseline anomalies. In
Sec.~\ref{sec:nsi} we derive limits on the presence of non-standard
neutrino-matter interactions.

\section{Analysis in the framework of three-neutrino mixing}
\label{sec:3numixing}

The wealth of data listed in the introduction can be consistently
described by assuming mixing among the three known neutrinos ($\nu_e$,
$\nu_\mu$, $\nu_\tau$), which can be expressed as quantum
superpositions of three massive states $\nu_i$ ($i=1,2,3$) with masses
$m_i$.  As explained in the previous section this implies the presence
of a leptonic mixing matrix in the weak charged current interactions
which can be parametrized as~\cite{Agashe:2014kda}:
\begin{equation}
  \label{eq:matrix}
  U =
  \begin{pmatrix}
    c_{12} c_{13}
    & s_{12} c_{13}
    & s_{13} e^{-i\delta_\text{CP}}
    \\
    - s_{12} c_{23} - c_{12} s_{13} s_{23} e^{i\delta_\text{CP}}
    & \hphantom{+} c_{12} c_{23} - s_{12} s_{13} s_{23}
    e^{i\delta_\text{CP}}
    & c_{13} s_{23}
    \\
    \hphantom{+} s_{12} s_{23} - c_{12} s_{13} c_{23} e^{i\delta_\text{CP}}
    & - c_{12} s_{23} - s_{12} s_{13} c_{23} e^{i\delta_\text{CP}}
    & c_{13} c_{23}
  \end{pmatrix}
  \begin{pmatrix}
    e^{i\alpha_1} & 0 & 0 \\
    0& e^{i\alpha_2} & 0 \\
    0 & 0 & 1
  \end{pmatrix}
\end{equation}
where $c_{ij} \equiv \cos\theta_{ij}$ and $s_{ij} \equiv
\sin\theta_{ij}$.  In addition to the Dirac-type phase
$\delta_\text{CP}$, analogous to that of the quark sector, there are
two extra phases $\alpha_1$, $\alpha_2$ associated to a possible
Majorana character of neutrinos. Such phases, however, are not
relevant for neutrino oscillations.

In this convention, disappearance of solar $\nu_e$'s and long baseline
reactor $\bar\nu_e$'s proceeds dominantly via oscillations with
wavelength $\propto E / \Dmq_{21}$ ($\Dmq_{ij} \equiv m_i^2 - m_j^2$
and $\Dmq_{21} \geq 0$ by convention) and amplitudes controlled by
$\theta_{12}$, while disappearance of atmospheric and LBL accelerator
$\nu_\mu$'s proceeds dominantly via oscillations with wavelength
$\propto E/|\Dmq_{31}| \ll E/\Dmq_{21}$ and amplitudes controlled by
$\theta_{23}$. The angle $\theta_{13}$ controls the amplitude of
oscillations involving $\nu_e$ flavor with $E/|\Dmq_{31}|$
wavelengths.  Given the observed hierarchy between the solar and
atmospheric wavelengths there are two possible non-equivalent
orderings for the mass eigenvalues, which are conventionally chosen as
\begin{align}
  \label{eq:normal}
  \Dmq_{21} &\ll \hphantom{+} (\Dmq_{32} \simeq \Dmq_{31} > 0) \,;
  \\
  \label{eq:inverted}
  \Dmq_{21} &\ll -(\Dmq_{31} \simeq \Dmq_{32} < 0) \,,
\end{align}
As it is customary we refer to the first option,
Eq.~\eqref{eq:normal}, as Normal Ordering (NO), and to the second one,
Eq.~\eqref{eq:inverted}, as Inverted Ordering (IO); in this form they
correspond to the two possible choices of the sign of $\Dmq_{31}$. In
this convention the angles $\theta_{ij}$ can be taken without loss of
generality to lie in the first quadrant, $\theta_{ij} \in [0, \pi/2]$,
and the CP phase $\delta_\text{CP} \in [0, 2\pi]$.
In the following we adopt the (arbitrary) convention of reporting
results for $\Dmq_{31}$ for NO and $\Dmq_{32}$ for IO, \textit{i.e.},
we always use the one which has the larger absolute value. Sometimes
we will generically denote such quantity as $\Dmq_{3\ell}$, with
$\ell=1$ for NO and $\ell=2$ for IO.

\begin{table}[t]\centering
  \begin{tabular}{l|l|l}
    \hline\hline
    Experiment
    & Dominant & Important
    \\
    \hline
    Solar Experiments
    & $\theta_{12}$ & $\Dmq_{21}$, $\theta_{13}$
    \\
    Reactor LBL (KamLAND) & $\Dmq_{21}$
    & $\theta_{12}$, $\theta_{13}$
    \\
    Reactor MBL (Daya-Bay, Reno, D-Chooz)
    & $\theta_{13}$ & $|\Dmq_{3\ell}|$
    \\
    Atmospheric Experiments
    & $\theta_{23}$ & $|\Dmq_{3\ell}|$, $\theta_{13}$, $\delta_\text{CP}$
    \\
    Accelerator LBL $\nu_\mu$ Disapp (Minos, NO$\nu$A, T2K)
    & $|\Dmq_{3\ell }|$, $\theta_{23}$ & ~
    \\
    Accelerator LBL $\nu_e$ App (Minos, NO$\nu$A, T2K)
    & $\delta_\text{CP}$ & $\theta_{13}$, $\theta_{23}$, $\sign(\Dmq_{3\ell})$
    \\
    \hline\hline
  \end{tabular}
  \caption{Experiments contributing to the present determination of
    the oscillation parameters.}
  \label{tab:expe}
\end{table}

In summary, the $3\nu$ oscillation analysis of the existing data
involves six parameters: 2 mass differences (one of which can be
positive or negative), 3 mixing angles, and the CP phase
$\delta_\text{CP}$.  For the sake of clarity we summarize in
Table~\ref{tab:expe} which experiment contribute dominantly to the
present determination of the different parameters.

\begin{figure}[t]\centering
  \includegraphics[width=0.8\textwidth]{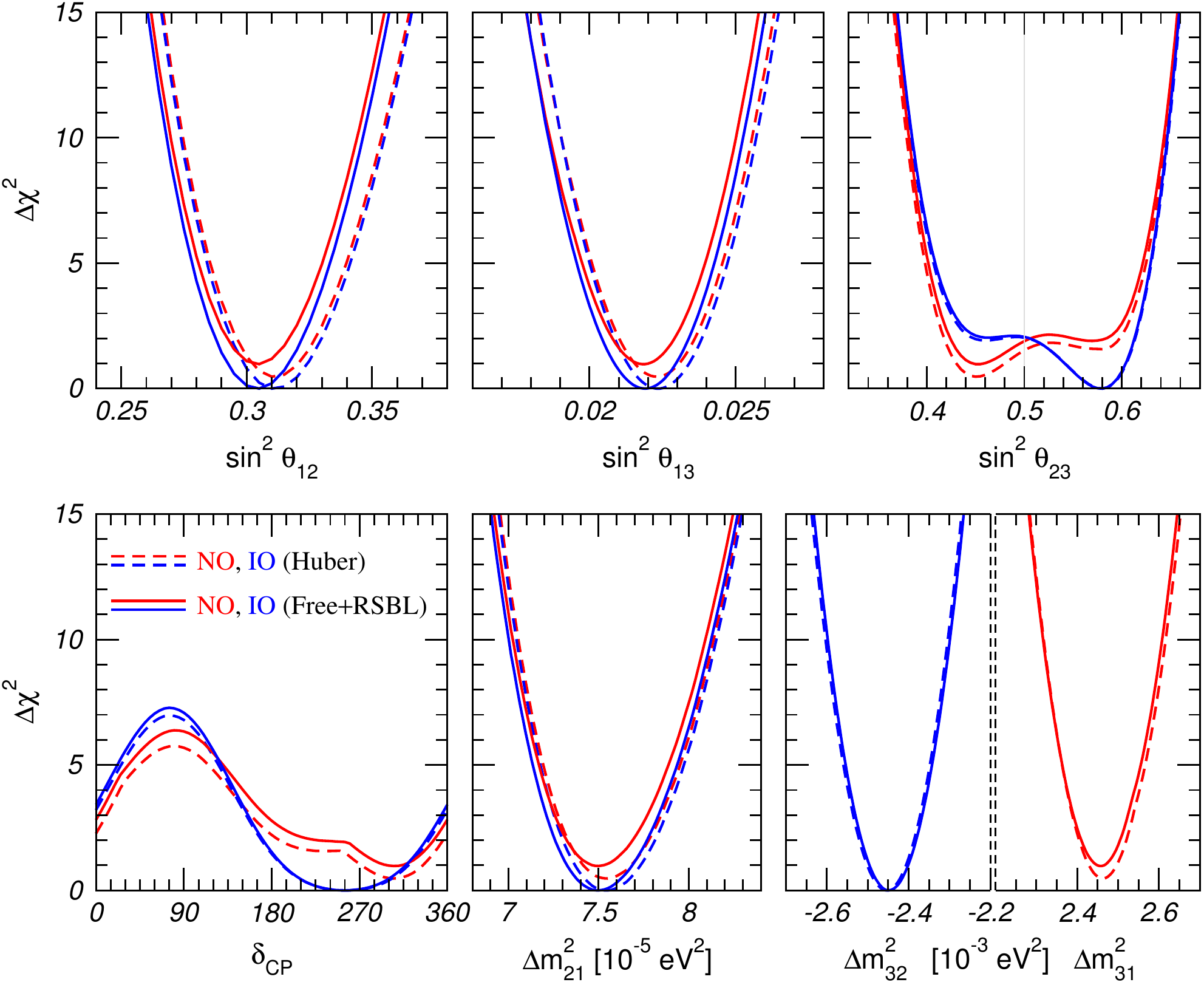}
  \caption{Global $3\nu$ oscillation analysis.  The red (blue) curves
    are for Normal (Inverted) Ordering.  For solid curves the
    normalization of reactor fluxes is left free and data from
    short-baseline (less than 100 m) reactor experiments are included.
    For dashed curves short-baseline data are not included but reactor
    fluxes as predicted in~\cite{Huber:2011wv} are assumed. Note that
    as atmospheric mass-squared splitting we use $\Dmq_{31}$ for NO
    and $\Dmq_{32}$ for IO. Figure similar to Fig.~2 in
    Ref.~\cite{Gonzalez-Garcia:2014bfa}.}
  \label{fig:3nu-chisq}
\end{figure}

The consistent determination of these leptonic parameters requires a
global analysis of the data described above. Such global fits are
presently performed by a few phenomenological
groups~\cite{Capozzi:2013csa, Forero:2014bxa,
  Gonzalez-Garcia:2014bfa}; here we summarize the results from
Ref.~\cite{Gonzalez-Garcia:2014bfa, nufit-2.0}.  We show in
Fig.~\ref{fig:3nu-chisq} the one-dimensional projections of the
$\Delta\chi^2$ of the global analysis as a function of each of the six
parameters.  The corresponding best fit values and the derived ranges
for the six parameters at the $1\sigma$ ($3\sigma$) level are given in
Tab.~\ref{tab:results}.  For each parameter the curves and ranges are
obtained after marginalizing with respect to the other five
parameters. The results in the table are shown for three scenarios.
In the first and second columns we assume that the ordering of the
neutrino mass states is known ``a priori'' to be Normal or Inverted,
respectively, so the ranges of all parameters are defined with respect
to the minimum in the given scenario.  In the third column we make no
assumptions on the ordering, so in this case the parameter ranges are
defined with respect to the global minimum (which corresponds to
Inverted Ordering) and are obtained marginalizing also over the
ordering. For this third case we only give the $3\sigma$ intervals. Of
course in this case the range of $\Dmq_{3\ell}$ is composed of two
disconnected intervals, one one containing the absolute minimum (IO)
and the other the secondary local minimum (NO).

As already mentioned, all the data described above can be consistently
interpreted as oscillations of the three known active neutrinos. In
addition to these data, however, several anomalies at short baselines
(SBL) have been observed which cannot be explained as $3\nu$
oscillations, but could be interpreted as oscillations involving an
$\mathcal{O}(\text{eV})$ mass sterile state. They will be discussed in
detail in Sec.~\ref{sec:1evsterile}. For what concerns the analysis
presented here the only SBL effect which has to be taken into account
is the so called \textit{reactor anomaly}. It turns out that the most
recent reactor flux calculations~\cite{Mueller:2011nm, Huber:2011wv}
fall short at describing the results from reactor experiments at
baselines $\lesssim 100$~m, such as Bugey4~\cite{Declais:1994ma},
ROVNO4~\cite{Kuvshinnikov:1990ry}, Bugey3~\cite{Declais:1994su},
Krasnoyarsk~\cite{Vidyakin:1987ue, Vidyakin:1994ut},
ILL~\cite{Kwon:1981ua}, G\"osgen~\cite{Zacek:1986cu},
SRP~\cite{Greenwood:1996pb}, and ROVNO88~\cite{Afonin:1988gx}.  Such
reactor short-baseline experiments (RSBL) do not contribute to
oscillation physics in the $3\nu$ framework, but they play an
important role in constraining the unoscillated reactor neutrino flux
if they are used instead of the theoretically calculated reactor
fluxes.  Thus to account for the possible effect of the reactor
anomaly in the determined ranges of neutrino parameters we show the
results in Fig.~\ref{fig:3nu-chisq} for two extreme choices.  The
first option (labeled ``Free+RSBL'') is to leave the normalization of
reactor fluxes free and include the RSBL data. The second option
(labeled ``Huber'') is not to include short-baseline reactor data but
assume reactor fluxes and uncertainties as predicted
in~\cite{Huber:2011wv}.

\begin{table}[t]\centering
  \begin{tabular}{@{~}l|cc|cc|c@{~}}
    \hline\hline
    & \multicolumn{2}{c|}{Normal Ordering ($\Delta\chi^2=0.97$)}
    & \multicolumn{2}{c|}{Inverted Ordering (best fit)}
    & Any Ordering
    \\
    \hline
    & bfp $\pm 1\sigma$ & $3\sigma$ range
    & bfp $\pm 1\sigma$ & $3\sigma$ range
    & $3\sigma$ range
    \\
    \hline
    \rule{0pt}{4mm}\ignorespaces
    $\sin^2\theta_{12}$
    & $0.304_{-0.012}^{+0.013}$ & $0.270 \to 0.344$
    & $0.304_{-0.012}^{+0.013}$ & $0.270 \to 0.344$
    & $0.270 \to 0.344$
    \\[1mm]
    $\theta_{12}/^\circ$
    & $33.48_{-0.75}^{+0.78}$ & $31.29 \to 35.91$
    & $33.48_{-0.75}^{+0.78}$ & $31.29 \to 35.91$
    & $31.29 \to 35.91$
    \\[3mm]
    $\sin^2\theta_{23}$
    & $0.452_{-0.028}^{+0.052}$ & $0.382 \to 0.643$
    & $0.579_{-0.037}^{+0.025}$ & $0.389 \to 0.644$
    & $0.385 \to 0.644$
    \\[1mm]
    $\theta_{23}/^\circ$
    & $42.3_{-1.6}^{+3.0}$ & $38.2 \to 53.3$
    & $49.5_{-2.2}^{+1.5}$ & $38.6 \to 53.3$
    & $38.3 \to 53.3$
    \\[3mm]
    $\sin^2\theta_{13}$
    & $0.0218_{-0.0010}^{+0.0010}$ & $0.0186 \to 0.0250$
    & $0.0219_{-0.0010}^{+0.0011}$ & $0.0188 \to 0.0251$
    & $0.0188 \to 0.0251$
    \\[1mm]
    $\theta_{13}/^\circ$
    & $8.50_{-0.21}^{+0.20}$ & $7.85 \to 9.10$
    & $8.51_{-0.21}^{+0.20}$ & $7.87 \to 9.11$
    & $7.87 \to 9.11$
    \\[3mm]
    $\delta_\text{CP}/^\circ$
    & $306_{-70}^{+39}$ & $\hphantom{00}0 \to 360$
    & $254_{-62}^{+63}$ & $\hphantom{00}0 \to 360$
    & $\hphantom{00}0 \to 360$
    \\[3mm]
    $\frac{\Dmq_{21}}{10^{-5}~\eVq}$
    & $7.50_{-0.17}^{+0.19}$ & $7.02 \to 8.09$
    & $7.50_{-0.17}^{+0.19}$ & $7.02 \to 8.09$
    & $7.02 \to 8.09$
    \\[3mm]
    $\frac{\Dmq_{3\ell}}{10^{-3}~\eVq}$
    & $+2.457_{-0.047}^{+0.047}$ & $+2.317 \to +2.607$
    & $-2.449_{-0.047}^{+0.048}$ & $-2.590 \to -2.307$
    & $\begin{bmatrix}
      +2.325 \to +2.599\\[-1pt]
      -2.590 \to -2.307
    \end{bmatrix}$
    \\[3mm]
    \hline\hline
  \end{tabular}
  \caption{Three-flavor oscillation parameters from our fit to global
    data after the NOW~2014 conference~\cite{Gonzalez-Garcia:2014bfa}.
    The results are presented for the ``Free Fluxes + RSBL'' in which
    reactor fluxes have been left free in the fit and short baseline
    reactor data (RSBL) with $L \lesssim 100$~m are included. The
    numbers in the 1st (2nd) column are obtained assuming NO (IO),
    \textit{i.e.}, relative to the respective local minimum, whereas
    in the 3rd column we minimize also with respect to the
    ordering. Note that $\Dmq_{3\ell} \equiv \Dmq_{31} > 0$ for NO and
    $\Dmq_{3\ell} \equiv \Dmq_{32} < 0$ for IO.}
  \label{tab:results}
\end{table}

From the results in the figure and table we conclude that:
\begin{enumerate}
\item if we define the $3\sigma$ relative precision of a parameter by
  $2(x^\text{up} - x^\text{low}) / (x^\text{up} + x^\text{low})$,
  where $x^\text{up}$ ($x^\text{low}$) is the upper (lower) bound on a
  parameter $x$ at the $3\sigma$ level, from the numbers in the table
  we find $3\sigma$ relative precision of 14\% ($\theta_{12}$), 32\%
  ($\theta_{23}$), 15\% ($\theta_{13}$), 14\% ($\Dmq_{21}$) and 11\%
  ($|\Dmq_{3\ell}|$) for the various oscillation parameters;

\item for either choice of the reactor fluxes the global best fit
  corresponds to IO with $\sin^2\theta_{23} > 0.5$, while the second
  local minima is for NO with $\sin^2\theta_{23} < 0.5$;

\item the statistical significance of the preference for Inverted
  versus Normal ordering is quite small, $\Delta\chi^2 \lesssim 1$;

\item the present global analysis disfavors $\theta_{13}=0$ with a
  $\Delta\chi^2 \approx 500$. Such impressive result is mostly driven
  by the reactor data from Daya Bay, with secondary contributions from
  RENO and Double Chooz;

\item the uncertainty on $\theta_{13}$ associated with the choice of
  reactor fluxes is at the level of $0.5\sigma$ in the global
  analysis. This is so because the most precise results from Daya Bay
  and RENO are independent of the reactor flux normalization;

\item a non-maximal value of the $\theta_{23}$ mixing is slightly
  favored, at the level of $\sim 1.4\sigma$ for Inverted Ordering at
  of $\sim 1.0\sigma$ for Normal Ordering;

\item the statistical significance of the preference of the fit for
  the second (first) octant of $\theta_{23}$ is $\leq 1.4\sigma$
  ($\leq 1.0\sigma$) for IO (NO);

\item the best fit for $\delta_\text{CP}$ for all analyses and
  orderings occurs for $\delta_\text{CP} \simeq 3\pi/2$, and values
  around $\pi/2$ are disfavored with $\Delta\chi^2 \simeq 6$.
  Assigning a confidence level to this $\Delta\chi^2$ is non-trivial,
  due to the non-Gaussian behavior of the involved $\chi^2$ function,
  see Refs.~\cite{Gonzalez-Garcia:2014bfa, Elevant:2015ska} for
  discussions and a Monte Carlo studies.
\end{enumerate}
These results are robust with respect to changes in the statistical
interpretation. The Bayesian analysis performed
in~\cite{Bergstrom:2015rba} leads to quantitatively very similar
results.

From this global analysis one can also derive the $3\sigma$ ranges on
the magnitude of the elements of the leptonic mixing matrix to be:
\begin{equation}
  \label{eq:umatrix}
  |U| = \begin{pmatrix}
    0.801 \to 0.845 &\qquad
    0.514 \to 0.580 &\qquad
    0.137 \to 0.158
    \\
    0.225 \to 0.517 &\qquad
    0.441 \to 0.699 &\qquad
    0.614 \to 0.793
    \\
    0.246 \to 0.529 &\qquad
    0.464 \to 0.713 &\qquad
    0.590 \to 0.776
  \end{pmatrix} \,.
\end{equation}

\begin{figure}[t]\centering
  \includegraphics[width=0.7\textwidth]{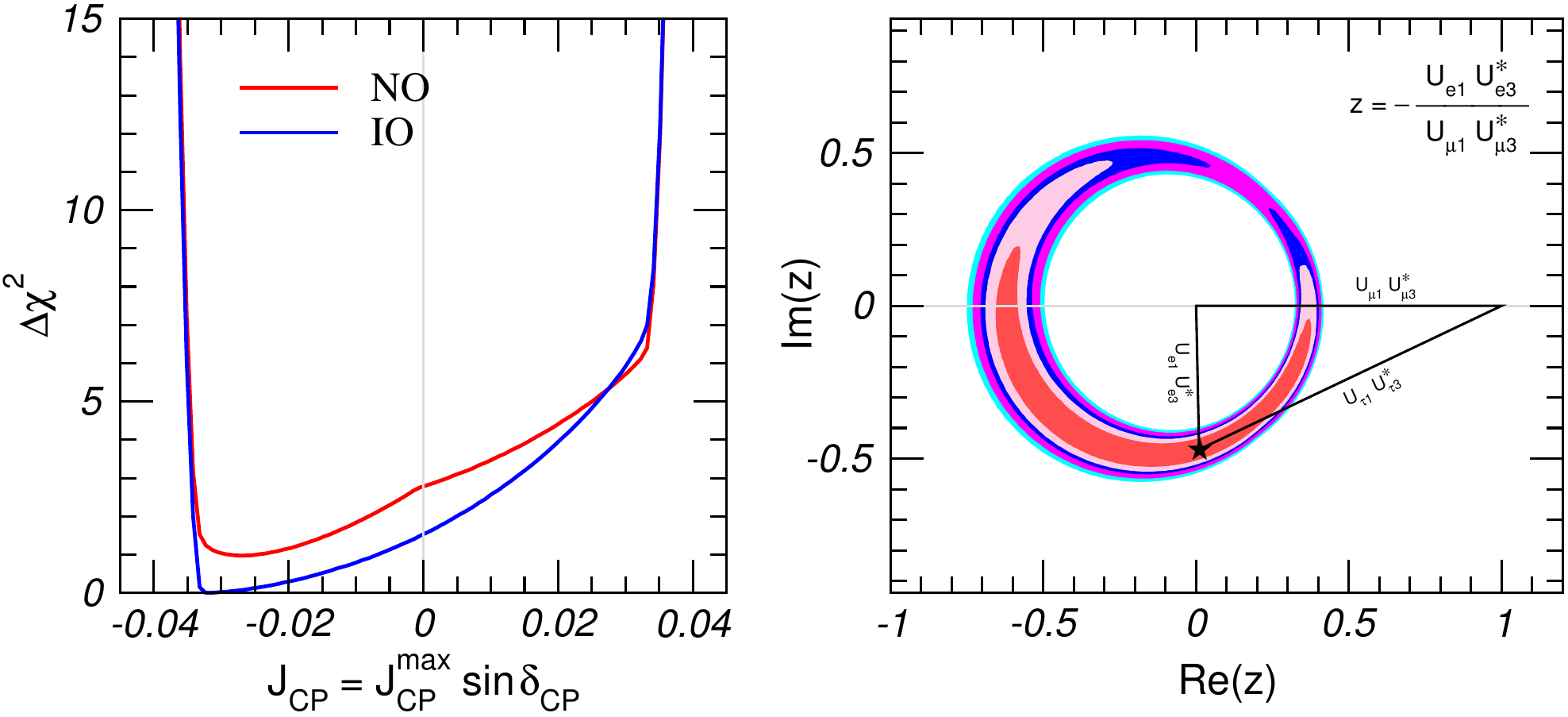}
  \caption{Left:dependence of the global $\Delta\chi^2$ function on
    the Jarlskog invariant. The red (blue) curves are for NO (IO).
    Right: leptonic unitarity triangle. After scaling and rotating so
    that two of its vertices always coincide with $(0,0)$ and $(1,0)$
    we plot the $1\sigma$, 90\%, $2\sigma$, 99\%, $3\sigma$ CL (2~dof)
    allowed regions of the third vertex.}
  \label{fig:3nu-viola}
\end{figure}

The present status of the determination of leptonic CP violation is
further illustrated in Fig.~\ref{fig:3nu-viola}. On the left panel we
show the dependence of $\Delta\chi^2$ of the global analysis on the
Jarlskog invariant which gives a convention-independent measure of CP
violation~\cite{Jarlskog:1985ht}, defined as:
\begin{equation}
  \Im\big[ U_{\alpha i} U_{\alpha j}^* U_{\beta i}^* U_{\beta j} \big]
  \equiv
  \cos\theta_{12} \sin\theta_{12} \cos\theta_{23} \sin\theta_{23}
  \cos^2\theta_{13} \sin\theta_{13} \, \sin\delta_\text{CP}
  \equiv
  J_\text{CP}^\text{max} \, \sin\delta_\text{CP}
\end{equation}
where we have used the parametrization in Eq.~\eqref{eq:matrix}.  Thus
the determination of the mixing angles yields at present a maximum
allowed CP violation
\begin{equation}
  \label{eq:jmax}
  J_\text{CP}^\text{max} = 0.0329 \pm 0.0009 \; (\mathrel{\pm} 0.0027)
\end{equation}
at $1\sigma$ ($3\sigma$) for both orderings.  The preference of the
present data for non-zero $\delta_\text{CP}$ implies a best fit
$J_\text{CP}^\text{best} = -0.032$, which is favored over CP
conservation at the $\sim 1.2\sigma$ level. These numbers can be
compared with the size of the Jarlskog invariant in the quark sector,
which is determined to be $J_\text{CP}^\text{quarks} =
(3.06_{-0.20}^{+0.21}) \times 10^{-5}$~\cite{Agashe:2014kda}.

In the right panel of Fig.~\ref{fig:3nu-viola} we recast the allowed
regions for the leptonic mixing matrix in terms of one leptonic
unitarity triangle. Since in the analysis $U$ is unitary by
construction, any given pair of rows or columns can be used to define
a triangle in the complex plane. In the figure we show the triangle
corresponding to the unitarity conditions on the first and third
columns which is the leptonic analogous to the one usually showed for
the quark sector.  In this figure the absence of CP violation implies
a flat triangle, \textit{i.e.}, $\Im(z) = 0$.  As can be seen, the
horizontal axis marginally crosses the $1\sigma$ allowed region, which
for 2~dof corresponds to $\Delta\chi^2 \simeq 2.3$. This is consistent
with the present preference for CP violation, $\chi^2(J_\text{CP} = 0)
- \chi^2(J_\text{CP}~\text{free}) = 1.5$.  A detailed discussion of
the status of the CP phase from present data can be found in
Ref.~\cite{Elevant:2015ska}.

\section{Absolute neutrino mass measurements}
\label{sec:absmass}

\begin{figure}[t]\centering
  \includegraphics[width=0.7\textwidth]{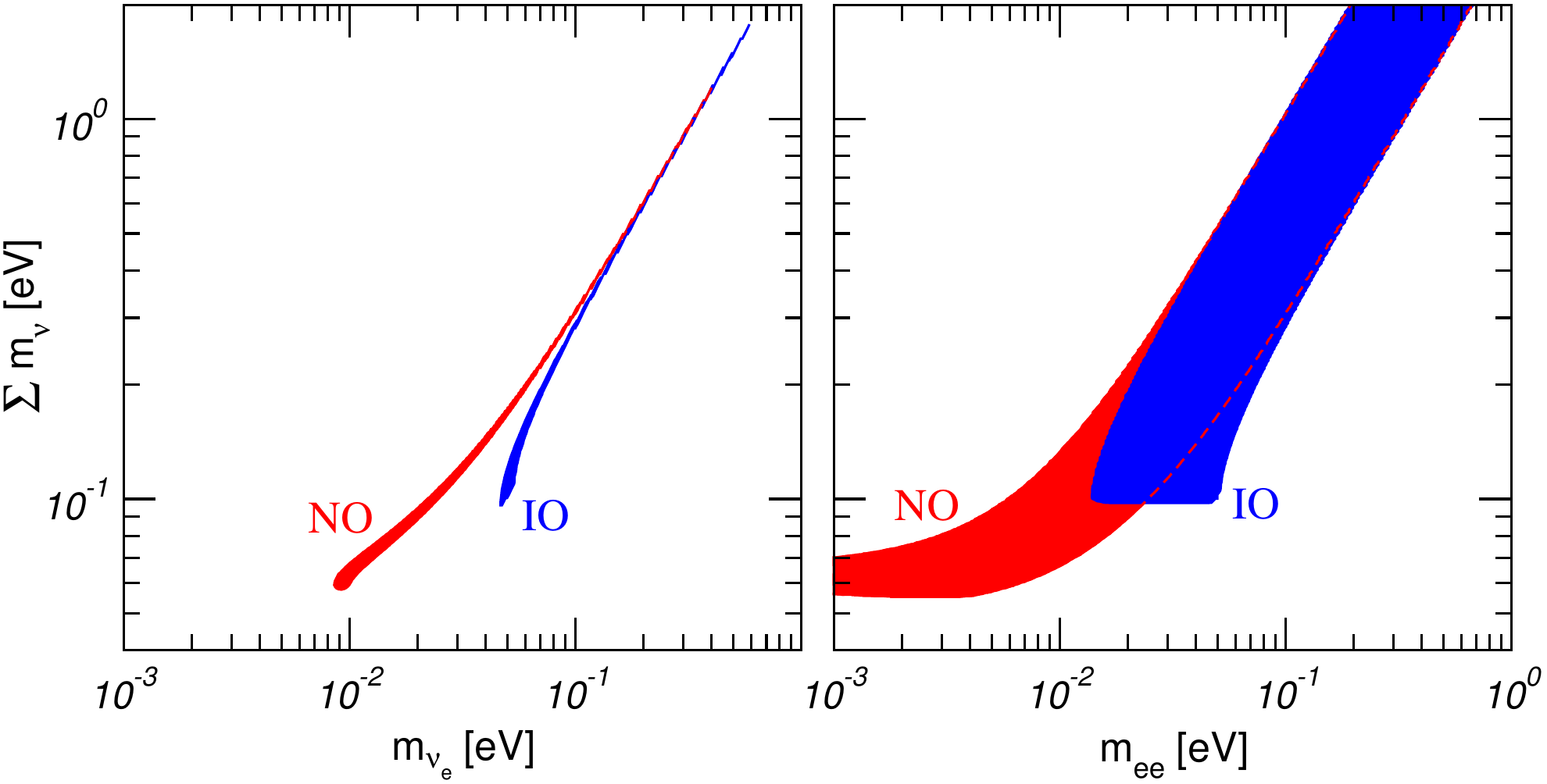}
  \caption{95\% allowed regions (for 2 dof) in the planes
    ($m_{\nu_e}$, $\sum m_\nu$) and ($m_{ee}$, $\sum m_\nu$) obtain
    from projecting the results of the global analysis of oscillation
    data.}
  \label{fig:3nu-mbeta}
\end{figure}

Oscillation experiments provide information on the mass-squared
splittings $\Dmq_{ij}$ and on the leptonic mixing angles $U_{ij}$, but
they are insensitive to the absolute mass scale for the neutrinos.  Of
course, the results of an oscillation experiment do provide a lower
bound on the heavier mass in $\Dmq_{ij}$, $|m_i| \geq
\sqrt{\Dmq_{ij}}$ for $\Dmq_{ij} > 0$, but there is no upper bound on
this mass. In particular, the corresponding neutrinos could be
approximately degenerate at a mass scale that is much higher than
$\sqrt{\Dmq_{ij}}$.  Moreover, there is neither an upper nor a lower
bound on the lighter mass $m_j$.

Information on the neutrino masses, rather than mass differences, can
be extracted from kinematic studies of reactions in which a neutrino
or an anti-neutrino is involved. In the presence of mixing the most
relevant constraint comes from the study of the end point ($E \sim
E_0$) of the electron spectrum in Tritium beta decay $\Nuc{3}{H} \to
\Nuc{3}{He} + e^- + \bar\nu_e$. This spectrum can be effectively
described by a single parameter, $m_{\nu_e}$, if for all neutrino
states $E_0 - E \gg m_i$. In this case:
\begin{equation}
  \frac{dN}{dE}
  \simeq R(E) \sum_i |U_{ei}|^2
  \sqrt{(E_0 - E)^2 - m_{\nu_e}^2} \,,
\end{equation}
where $R(E)$ contains all the $m_\nu$-independent factors, and
\begin{equation}
  \label{eq:mbeta}
  m^2_{\nu_e} = \frac{\sum_i m^2_i |U_{ei}|^2}{\sum_i |U_{ei}|^2}
  = \sum_i m^2_i |U_{ei}|^2
  = c_{13}^2 c_{12}^2 m_1^2
  + c_{13}^2 s_{12}^2 m_2^2+s_{13}^2 m_3^2 \,,
\end{equation}
where the second equality holds if unitarity is assumed.  At present
we only have an upper bound, $m_{\nu_e} \leq 2.2$~eV at 95\%
CL~\cite{Bonn:2001tw}, which is expected to be superseded soon by
KATRIN~\cite{Osipowicz:2001sq} with about one order of magnitude
improvement in sensitivity.

Direct information on neutrino masses can also be obtained from
neutrinoless double beta decay $(A,Z) \to (A,Z+2) + e^- + e^-$.  This
process violates lepton number by two units, hence in order to induce
the $0\nu\beta\beta$ decay $\nu$'s must Majorana particles. In
particular, for the case in which the only effective lepton number
violation at low energies is induced by the Majorana mass term for the
neutrinos, the rate of $0\nu\beta\beta$ decay is proportional to the
\emph{effective Majorana mass of $\nu_e$}:
\begin{equation}
  m_{ee}
  = \Big| \sum_i m_i U_{ei}^2 \Big|
  = \Big| m_1 c_{13}^2 c_{12}^2 e^{i 2\alpha_1} +
  m_2 c_{13}^2 s_{12}^2 e^{i 2\alpha_2} +
  m_3 s_{13}^2 e^{-i 2\delta_\text{CP}} \Big|
\end{equation}
which, unlike Eq.~\eqref{eq:mbeta}, depends also on the three CP
violating phases.
Recent searches carried out with \Nuc{76}{Ge} (GERDA
experiment~\cite{Agostini:2013mzu}) and \Nuc{136}{Xe}
(KamLAND-Zen~\cite{Gando:2012zm} and EXO-200~\cite{Albert:2014awa}
experiments) have established the lifetime of this decay to be longer
than $10^{25}$ yr, corresponding to a limit on the neutrino mass of
$m_{ee} \leq 0.2-0.4$ eV at 90\% CL.  A series of new experiments is
planned with sensitivity of up to $m_{ee} \sim
0.01$~eV~\cite{GomezCadenas:2011it}.

Neutrino masses have also interesting cosmological effects. In
general, cosmological data mostly give information on the sum of the
neutrino masses, $\sum_i m_{i}$, while they have very little to say on
their mixing structure and on the ordering of the mass states.

Correlated information on these three probes of the neutrino mass
scale can be obtained by mapping the results from the global analysis
of oscillations presented previously. We show in Fig.~\ref{fig:3nu-mbeta}
the present status of this exercise.  The relatively large width of
the regions in the right panel are due to the unknown Majorana
phases. Thus from a positive determination of two of these probes
information can be obtained on the value of the Majorana phases and/or
the mass ordering.

\section{Sterile neutrinos at the eV scale}
\label{sec:1evsterile}

Besides the huge success of three-flavour oscillations described in
Sec.~\ref{sec:3numixing} there are some anomalies which cannot be
explained within the $3\nu$ framework and which might point towards
the existence of additional neutrino flavors (so-called sterile
neutrinos) with masses at the eV scale:
\begin{itemize}
\item the LSND experiment~\cite{Aguilar:2001ty} reports evidence for
  $\bar\nu_\mu\to\bar\nu_e$ transitions with $E/L \sim 1~\eVq$, where
  $E$ and $L$ are the neutrino energy and the distance between source
  and detector, respectively;

\item this effect is also searched for by the MiniBooNE
  experiment~\cite{AguilarArevalo:2012va}, which reports a yet
  unexplained event excess in the low-energy region of the electron
  neutrino and anti-neutrino event spectra. No significant excess is
  found at higher neutrino energies.  Interpreting the data in terms
  of oscillations, parameter values consistent with the ones from LSND
  are obtained;

\item radioactive source experiments at the Gallium solar neutrino
  experiments SAGE and GALLEX have obtained an event rate which is
  somewhat lower than expected. This effect can be explained by the
  hypothesis of $\nu_e$ disappearance due to oscillations with $\Dmq
  \gtrsim 1~\eVq$ (``Gallium anomaly'')~\cite{Acero:2007su,
    Giunti:2010zu};

\item state-of-the-art calculations of the neutrino flux emitted by
  nuclear reactors~\cite{Mueller:2011nm, Huber:2011wv} predict a
  neutrino rate which is a few percent higher than observed in
  short-baseline ($L \lesssim 100$~m) reactor experiments. A decreased
  rate at those distances can be explained by assuming $\bar\nu_e$
  disappearance due to oscillations with $\Dmq \sim 1~\eVq$ (``reactor
  anomaly'')~\cite{Mention:2011rk}.
\end{itemize}

Here we report the results of a global analysis from
Ref.~\cite{Kopp:2013vaa} of those data under the hypothesis of
additional neutrino species at the eV scale (see~\cite{Conrad:2012qt,
  Giunti:2013aea} for similar analyses). We introduce a neutrino
state, $\nu_4$, with a mass-squared difference $\Dmq_{41}$ much larger
than $|\Dmq_{31}|$. This situation is called ``3+1 mass scheme''. In
this case the oscillation probabilities for experiments exploring the
range $E/L \sim 1~\eVq$ are rather simple:
\begin{equation}
  P_{\alpha\alpha} = 1 - \sin^22\theta_{\alpha\alpha} \sin^2\Delta \,,
  \qquad
  P_{\mu e}  = \sin^22\theta_{\mu e} \sin^2\Delta\,,
\end{equation}
where $\Delta \equiv \Dmq_{41} L / 4E$ and the effective mixing angles
are defined as
\begin{equation}
  \sin^22\theta_{\alpha\alpha}  \equiv 4 |U_{\alpha 4}|^2(1 - |U_{ \alpha 4}|^2) \,,
  \qquad
  \sin^22\theta_{\mu e} \equiv 4 |U_{\mu 4}|^2 |U_{e 4}|^2 \,,
\end{equation}
with $\alpha = e,\mu$ and $U_{\alpha 4}$ are the elements of the
lepton mixing matrix describing the mixing of the 4th neutrino mass
state with the electron and muon flavour. There is no CP violation in
3+1 SBL oscillations and those relations apply for neutrinos as well
as antineutrinos. Neglecting quadratic terms in the mixing matrix
elements one has the following relation between the effective
amplitudes relevant for appearance and disappearance probabilities:
\begin{equation}
  \label{eq:3+1constr}
  4 \sin^22\theta_{\mu e} \approx
  \sin^22\theta_{e e}
  \sin^22\theta_{\mu\mu} \,.
\end{equation}
Dividing the relevant data into $\nu_e$ disappearance, $\nu_\mu$
disappearance, and $\nu_\mu\to\nu_e$ appearance searches, this
relation implies that the system is over-constrained. Indeed, as will
be discussed below, there is significant tension in the global data
and Eq.~\eqref{eq:3+1constr} makes it difficult to obtain a good fi to
all available data.

\begin{figure}[t]\centering
  \includegraphics[width=0.5\textwidth]{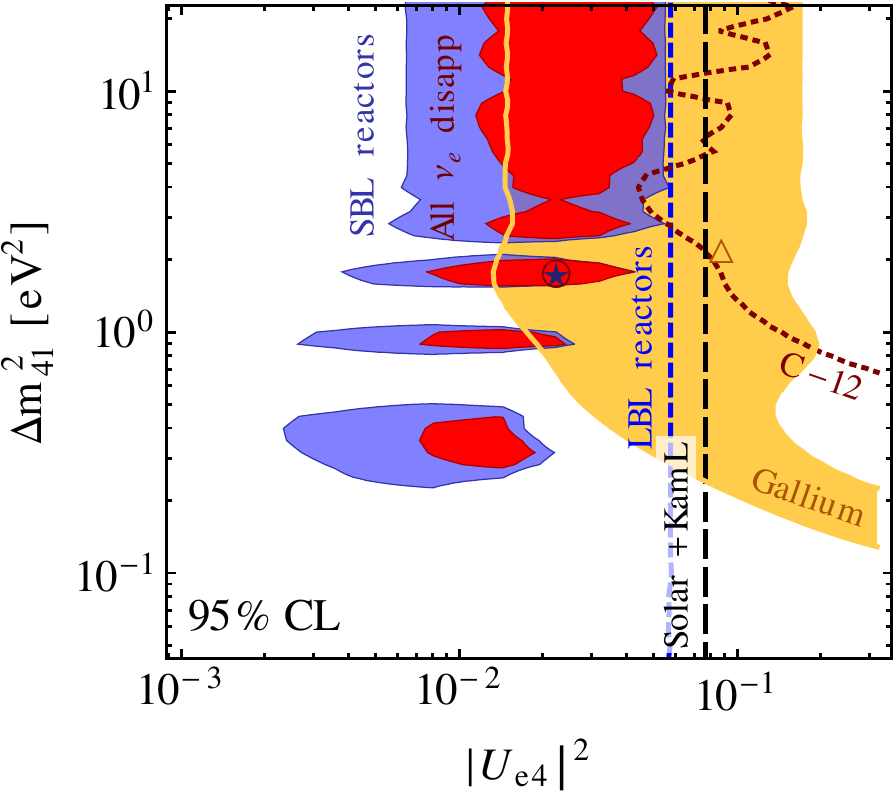}
  \caption{Allowed regions at 95\%~CL (2 dof) for 3+1 oscillations. We
    show SBL reactor data~\cite{Declais:1994ma, Kuvshinnikov:1990ry,
      Declais:1994su, Vidyakin:1987ue, Vidyakin:1994ut, Kwon:1981ua,
      Zacek:1986cu, Greenwood:1996pb, Afonin:1988gx} (blue shaded),
    Gallium radioactive source data~\cite{Hampel:1997fc,
      Kaether:2010ag, Abdurashitov:1998ne, Abdurashitov:2005tb}
    (orange shaded), $\nu_e$ disappearance constraints from
    $\nu_e$--$^{12}\text{C}$ scattering data from LSND and
    KARMEN~\cite{Auerbach:2001hz, Armbruster:1998uk} (dark red
    dotted), long-baseline reactor data from CHOOZ, Palo Verde,
    DoubleChooz, Daya Bay and RENO (blue short-dashed) and
    solar+KamLAND data (black long-dashed). The red shaded region is
    the combined region from all these $\nu_e$ and $\bar\nu_e$
    disappearance data sets. See Ref.~\protect\cite{Kopp:2013vaa} for
    details.}
  \label{fig:ste-disapp}
\end{figure}

We consider first the global data including SBL anomalies related to
$\bar\nu_e$ and $\nu_e$ disappearance (reactor and Gallium anomalies)
but ignoring for the time being the $\nu_\mu \to \nu_e$ and
$\bar\nu_\mu\to \bar\nu_e$ appearance anomalies (LSND and
MiniBooNE). In this case the relevant SBL phenomenology is determined
by the two parameters $\Dmq_{41}$ and $|U_{e4}|$. The allowed regions
for them is shown in Fig.~\ref{fig:ste-disapp}. We find that a
consistent region emerges (shown in red), not in conflict with any
other data. The best fit point occurs at $\sin^22\theta_{ee} = 0.09$
and $\Dmq_{41} = 1.78~\eVq$, and the no-oscillation hypothesis for the
eV-scale is excluded at $3.1\sigma$ ($\Delta\chi^2 = 12.9/2$~dof),
driven by the reactor and Gallium anomalies.
The $\theta_{13}$ determination is rather stable with respect to the
presence of sterile neutrinos, up to an ambiguity at the level of less
than $1\sigma$ (see also discussion in
section~\ref{sec:3numixing}). We note, however, that its
interpretation becomes slightly more complicated. For instance, using
a particular parametrization~\cite{Kopp:2013vaa} for the 3+1 scheme,
the relation between mixing matrix elements and mixing angles is
$|U_{e3}| = \cos\theta_{14} \sin\theta_{13}$ and $|U_{e4}| =
\sin\theta_{14}$. Hence, the one-to-one correspondence between
$|U_{e3}|$ and $\theta_{13}$ as in the three-flavor case is spoiled.

\begin{figure}[t]\centering
  \includegraphics[width=0.8\textwidth]{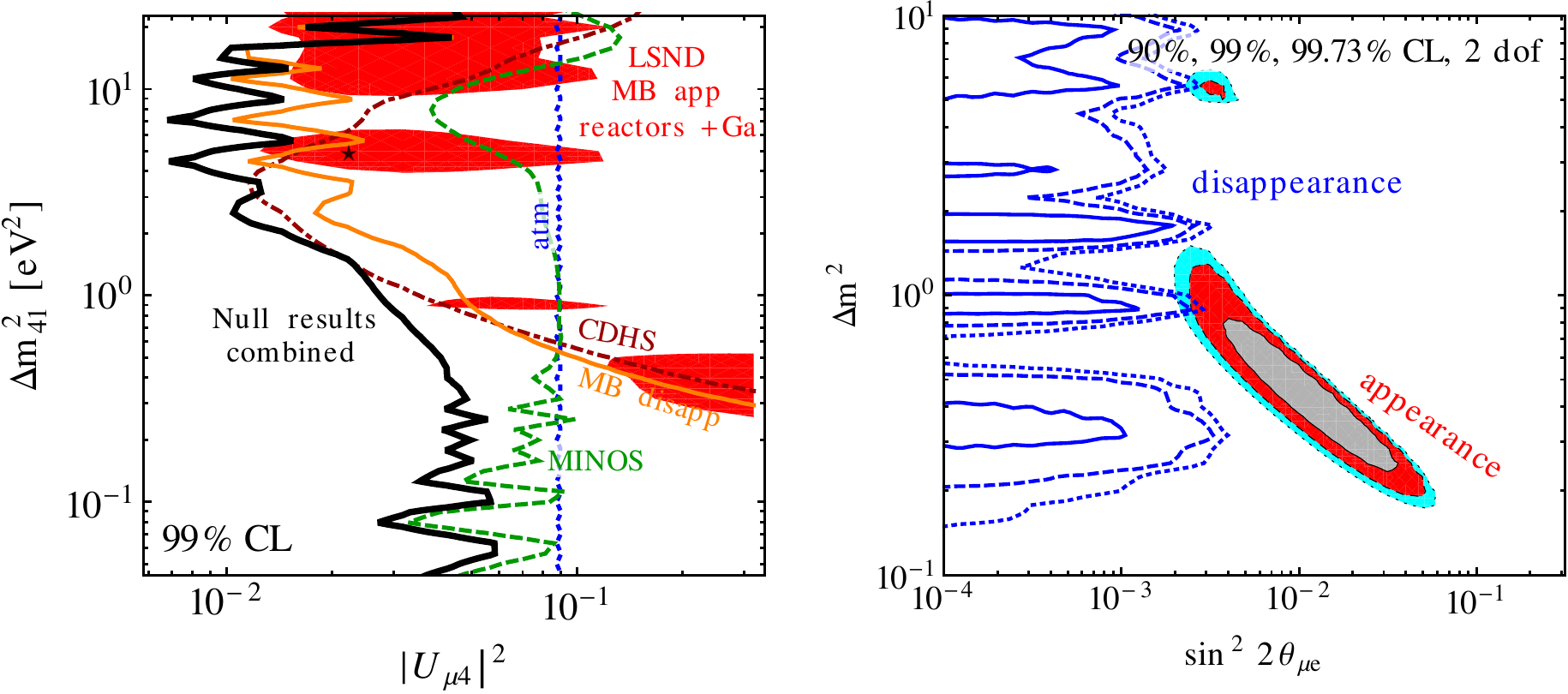}
  \caption{Left: Constraints in the plane of $|U_{\mu 4}|^2$ and
    $\Dmq_{41}$ at 99\%~CL (2~dof) from CDHS~\cite{Dydak:1983zq},
    atmospheric neutrinos~\cite{Wendell:2010md}, MiniBooNE
    disappearance~\cite{Cheng:2012yy}, MINOS CC and NC
    data~\cite{Adamson:2010wi, Adamson:2011ku}, and the combination of
    them. In red we show the region preferred by LSND and MiniBooNE
    appearance data combined with reactor and Gallium data, where for
    fixed $|U_{\mu 4}|^2$ we minimize with respect to $|U_{e4}|^2$.
    Right: Comparison of the parameter region preferred by appearance
    data (LSND~\cite{Aguilar:2001ty}, MiniBooNE appearance
    analysis~\cite{AguilarArevalo:2012va}, NOMAD~\cite{Astier:2003gs},
    KARMEN~\cite{Armbruster:2002mp}, ICARUS~\cite{Antonello:2012pq},
    E776~\cite{Borodovsky:1992pn}) to the exclusion limit from
    disappearance data (atmospheric, solar, reactors, Gallium, CDHS,
    MINOS, MiniBooNE disappearance, KARMEN and LSND
    $\nu_e-\Nuc{12}{C}$ scattering). See
    Ref.~\protect\cite{Kopp:2013vaa} for details.}
  \label{fig:ste-tension}
\end{figure}

We now address the question whether the hints for $\nu_e$
disappearance can be reconciled with the appearance hints from LSND
and MiniBooNE. As mentioned above, Eq.~\eqref{eq:3+1constr} links
those appearance signals to disappearance in the $\nu_e$ as well as
$\nu_\mu$ channels. Despite the possible signal in $\nu_e$
disappearance, so-far no positive signal has been observed in
$\nu_\mu$ disappearance and several experiments set bounds on the
relevant mixing parameter $|U_{\mu 4}|$, see
Fig.~\ref{fig:ste-tension}~(left). Hence, the combined limits on $\nu_\mu$ and
$\nu_e$ mixing with the eV-scale mass state lead to a tension between
appearance signals and disappearance data in the 3+1 scheme.
Such tension is illustrated for global data in the right panel of
Fig.~\ref{fig:ste-tension}, where we show the allowed region for all
appearance experiments, compared to the limit from disappearance
experiments in the plane of $\sin^22\theta_{\mu e}$ and
$\Dmq_{41}$. The preferred values of $\Dmq_{41}$ for disappearance
data come from the reactor and Gallium anomalies. The regions for
disappearance data, however, are not closed in this projection in the
parameter space and include $\sin^22\theta_{\mu e} = 4|U_{e 4} U_{\mu
  4}|^2 = 0$, which can always be achieved by letting $U_{\mu 4} \to
0$ due to the non-observation of any positive signal in SBL $\nu_\mu$
disappearance. The upper bound on $\sin^22\theta_{\mu e}$ from
disappearance emerges essentially as the product of the upper bounds
on $|U_{e 4}|$ and $|U_{\mu 4}|$ from $\nu_e$ and $\nu_\mu$
disappearance according to Eq.~\eqref{eq:3+1constr}.  We observe from
the plot the clear tension between those data sets, with only marginal
overlap regions at above 99\%~CL around $\Dmq_{41} \approx 0.9~\eVq$
and at 3$\sigma$ around $\Dmq_{41} \approx 6~\eVq$.  We find that the
global 3+1 fit leads to $\chi^2_\text{min}$/dof = 712/680 with a
p-value 19\%, whereas the so-called parameter goodness of fit (PG)
test~\cite{Maltoni:2003cu} indicates that appearance and disappearance
data are consistent with each other only with a p-value of about
$10^{-4}$.

A valid question to ask is whether the situation improves if more
neutrino states at the eV scale are introduced. Consider the
hypothesis of 2 states with eV scale mass splittings, $\nu_4$ and
$\nu_5$, which can be arranged either such that $\Dmq_{41}$ and
$\Dmq_{51}$ are both positive (``3+2'') and where one of them is
negative (``1+3+1''). The new qualitative feature in those 5-neutrino
schemes is CP violation at the $E/L \sim \eVq$
scale~\cite{Karagiorgi:2006jf, Maltoni:2007zf}, which introduces some
freedom in fitting neutrino versus anti-neutrino data from LSND and
MiniBooNE together. However, the main prediction from the 4-neutrino
case remains valid also for 5-neutrinos: a non-zero $\nu_\mu\to\nu_e$
appearance at SBL necessarly predicts SBL disappearance for $\nu_e$ as
well as $\nu_\mu$. Indeed, the tension between appearance and
disappearance data remains severe, and a PG analysis gives a
consistency below $10^{-4}$ for 3+2, whereas for 1+3+1 consistency at
the 2 permil level can be achieved~\cite{Kopp:2013vaa}.

In summary, several anomalies at the level of $3\sigma$ do not fit
into the three-flavour picture and might indicate additional neutrino
states at the eV scale. While a consistent fit can be obtained for
data on $\nu_e$ disappearance (reactor and Gallium anomalies) the
global data suffers from severe tension between appearance and
disappearance data, mostly due to the non-observation of $\nu_\mu$
disappearance at the \eVq\ scale. Finally we mention that additional
neutrino states with eV-like masses and sizeable mixings (as necessary
to explain the oscillation anomalies) have severe implications for
cosmology~\cite{Ade:2015xua, Bergstrom:2014fqa} and may lead to
observable effects in IceCube~\cite{Nunokawa:2003ep, ICste:jones,
  ICste:arguelles}.

\section{Matter potential: non-standard interactions}
\label{sec:nsi}

Neutrino oscillation experiments can also provide important
information on other neutrino properties beyond the SM. As an example
we briefly summarize here the results of the most up-to-date
determination of new physics in the matter effects in neutrino
propagation from the global analysis of neutrino oscillation
experiments from Ref.~\cite{Gonzalez-Garcia:2013usa}, to which we
refer the reader for details and related references.

In the three-flavor oscillation picture described above the neutrino
evolution equation along trajectory parametrized by coordinate $x$
reads ($\vec\nu=(\nu_e,\nu_\mu,\nu_\tau)^T$):
\begin{equation}
  i\frac{d}{dx} \vec\nu
  = (H_\text{vac} + H_\text{mat}) \, \vec\nu
  \quad \text{with} \quad
  H_\text{vac} = U D_\text{vac} U^\dagger \,,
  \quad
  D_\text{vac} = \frac{1}{2E_\nu} \diag(0, \Dmq_{21}, \Dmq_{31})
\end{equation}
while for antineutrinos the hamiltonian is $H^{\bar\nu} =
(H_\text{vac} - H_\text{mat})^*$.
In the Standard Model $H_\text{mat}$ is fully determined both in its
strength and flavor structure to be $H^\text{SM}_\text{mat} =\sqrt{2}
G_F N_e(r) \diag(1, 0, 0)$ for ordinary
matter~\cite{Wolfenstein:1977ue, Mikheev:1986gs}.  Generically
ordinary matter is composed by electrons ($e$), up-quarks ($u$) and
down-quark ($d$), thus in the most general case a non-standard matter
potential can be parametrized as:
\begin{equation}
  \label{eq:hmatNSI}
  H_\text{mat} = \sqrt{2} G_F N_e(r)
  \begin{pmatrix}
    1 & 0 & 0 \\
    0 & 0 & 0 \\
    0 & 0 & 0
  \end{pmatrix}
  + \sqrt{2} G_F \sum_{f=e,u,d} N_f(r)
  \begin{pmatrix}
    \Eps_{ee}^f & \Eps_{e\mu}^f & \Eps_{e\tau}^f
    \\
    \Eps_{e\mu}^{f*} & \Eps_{\mu\mu}^f & \Eps_{\mu\tau}^f
    \\
    \Eps_{e\tau}^{f*} & \Eps_{\mu\tau}^{f*} & \Eps_{\tau\tau}^f
  \end{pmatrix} \,.
\end{equation}
Since this matter term can be determined by oscillation experiments
only up to an overall multiple of the identity, without loss of
generality one can assume $\Eps_{\mu\mu}^f = 0$. With this, we have 8
additional parameters (for each $f$) since $\Eps_{ee}^f$ and
$\Eps_{\tau\tau}^f$ must be real whereas $\Eps_{e\mu}^f$,
$\Eps_{e\tau}^f$ and $\Eps_{\mu\tau}^f$ can be complex.

\begin{figure}[t]\centering
  \includegraphics[width=0.8\textwidth]{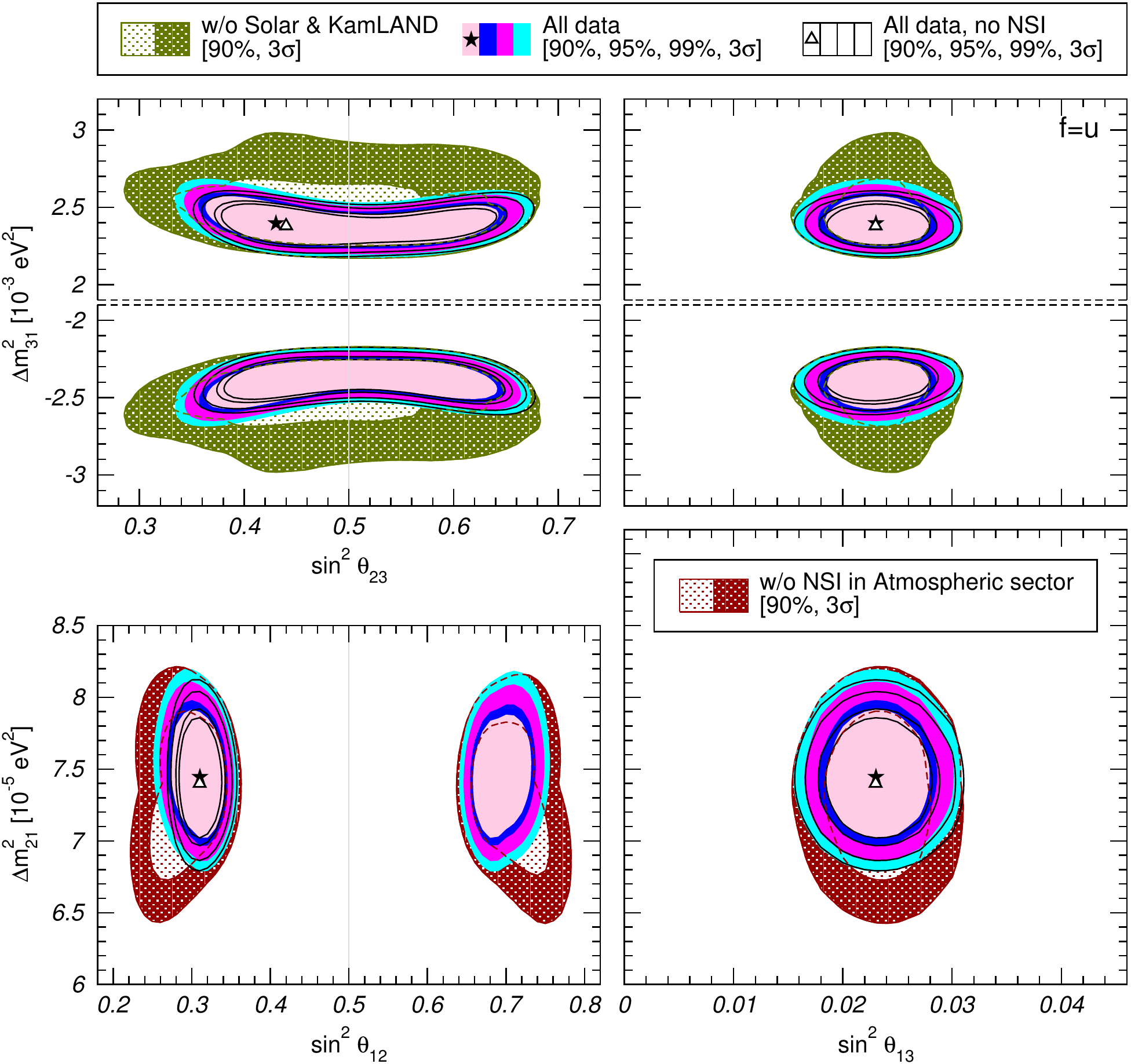}
  \caption{Two-dimensional projections of the 90\%, 95\%, 99\% and
    $3\sigma$ CL (2~dof) allowed regions of the oscillation parameters
    for $f=u$ after marginalizing over the matter potential parameters
    and the undisplayed oscillation parameters. The full regions and
    the star correspond to the global analysis including NSI, while
    the black-contour void regions and the triangle correspond to the
    analysis with the usual SM potential.  The green and red dotted
    areas show the 90\% and $3\sigma$ CL allowed regions from partial
    analyses where the effects of the non-standard matter potential
    have been neglected either in the solar+KamLAND (green) or in the
    atmospheric+LBL (red) sectors.}
  \label{fig:nsi-region}
\end{figure}

The theoretical framework for this parametrization of the matter
potential is provided by non-standard interactions (NSI) of neutrinos
with the matter particles.  They can be described by effective
four-fermion operators of the form
\begin{equation}
  \label{eq:def}
  \mathcal{L}_\text{NSI} =
  - 2\sqrt{2} G_F \Eps_{\alpha\beta}^{fP}
  (\bar\nu_{\alpha} \gamma^\mu \nu_{\beta})
  (\bar{f} \gamma_\mu P f) \,,
\end{equation}
where $f$ is a charged fermion, $P=(L,R)$ and
$\Eps_{\alpha\beta}^{fP}$ are dimensionless parameters encoding the
deviation from standard interactions.  NSI enter in neutrino
propagation only through the vector couplings so the induced matter
Hamiltonian takes the form Eq.~\eqref{eq:hmatNSI} with
$\Eps_{\alpha\beta}^f = \Eps_{\alpha\beta}^{fL} +
\Eps_{\alpha\beta}^{fR}$.

\begin{figure}[t]\centering
  \includegraphics[width=0.8\textwidth]{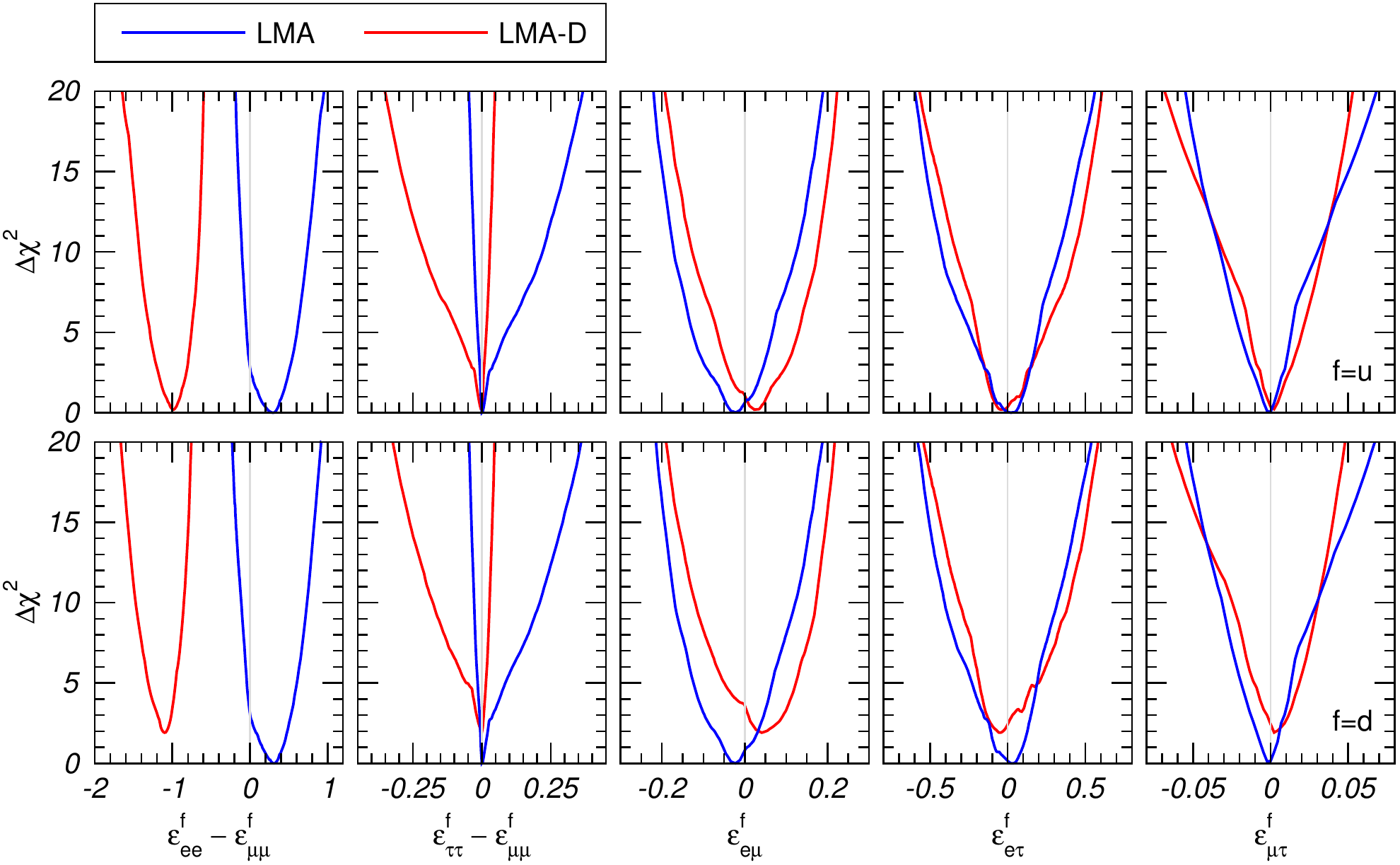}
  \caption{Dependence of the $\Delta\chi^2$ function for the global
    analysis of solar, atmospheric, reactor and LBL data on the NSI
    parameters $\Eps_{\alpha\beta}^f$ for $f=u$ (upper panels) and
    $f=d$ (lower panels), for both LMA and LMA-D regions and the two
    variants of the SNO analysis, as labeled in the figure.}
  \label{fig:nsi-chisq}
\end{figure}

We show in Figs.~\ref{fig:nsi-region} and \ref{fig:nsi-chisq} some
projections of the large parameter space in oscillation parameters and
on the NSI parameters (after marginalizing over all oscillation and
NSI undisplayed parameters) from a global analysis of oscillation data
in terms of $3\nu$ oscillations with general real matter potential
(with $\Dmq_{21}$ effects neglected in the analysis of ATM and LBL
experiments). From the figures we read the following:
\begin{itemize}
\item the determination of most the oscillation parameters discussed
  in the previous section is robust under the presence of NSI as large
  as allowed by the oscillation data itself with the exception of the
  octant of $\theta_{12}$;

\item a solution with $\theta_{12}>45 ^\circ$ (the ``so-called'' LMA-D
  solution~\cite{Miranda:2004nb}) still provides a good fit to the
  data, as can be seen in the lower-left panel in
  Fig.~\ref{fig:nsi-region}. Such solution requires large NSI, which
  nevertheless are fully compatible with the bounds from atmospheric
  and LBL oscillation data;

\item the analysis of solar and KamLAND data favours non-vanishing NSI
  to better fit the fact that neither the SNO nor SK4 low energy
  threshold analysis nor the $^8${B} measurement in Borexino seem to
  show evidence of the low energy turn-up of the spectrum predicted in
  the standard LMA MSW;

\item comparing the results in Fig.~\ref{fig:nsi-chisq} with the
  bounds on NSI derived in Refs.~\cite{Davidson:2003ha, Biggio:2009nt}
  from non-oscillation data we find that, with the possible exception
  of $\Eps_{e\mu}^{u,d}$, the global oscillation analysis presented
  here yields the most restrictive bounds on the \emph{vector} NSI
  parameters, in particular those involving $\tau$ flavour.
\end{itemize}
It is important to notice that in writing the phenomenological
Lagrangian in Eq.~\eqref{eq:def} one assumes that the new physics
which induces the NSI operators does not introduce new charge lepton
physics at tree level or that charge lepton effects are very
suppressed compared to those of neutrinos. This constraints the new
physics realizations of this scenario~\cite{Gavela:2008ra} and,
generically, the size of the NSI couplings which can be generated.

\section{Conclusions and outlook}

Thanks to the remarkable discoveries by neutrino oscillation
experiments, such us those awarded by 2015 Nobel prize, it is now an
established fact that neutrinos have mass and leptonic flavors are not
symmetries of Nature. These results represent, to this date, the only
laboratory evidence of physics beyond the Standard Model.

In this contribution we have summarized some results on the present
characterization of the low energy parametrization of the neutrino
properties as obtained from direct comparison with the data.  The
relevance of the work lies on the fact that the determination of the
flavour structure of the leptons at low energies, is, at this point,
our only source of information to understand the underlying new
dynamics and it is fundamental to ultimately establish the \emph{New
  Standard Model}.

\section*{Acknowledgments}

This work is supported by Spanish MINECO grants FPA2012-31880,
FPA2012-34694 and FPA2013-46570, by the Severo Ochoa program
SEV-2012-0249 of IFT, by the Maria de Maeztu program MDM-2014-0369 of
ICCUB, and consolider-ingenio 2010 grant CSD-2008-0037, by CUR
Generalitat de Catalunya grant 2014-SGR-104, by USA-NSF grant
PHY-13-16617, and by EU grant FP7 ITN INVISIBLES (Marie Curie Actions
PITN-GA-2011-289442).

\bibliography{biblio}
\bibliographystyle{elsarticle-num}

\end{document}